\documentclass[12pt]{article}
\usepackage[dvips]{color}
\usepackage{epsfig}
\usepackage{amsmath}
\usepackage{graphicx}

\begin{document}
\title{\bf  Tachyonic matter cosmology with exponential and hyperbolic potentials}
\author{{B. Pourhassan$^{a,}$\thanks{Email: b.pourhassan@du.ac.ir}, and J. Naji$^{b,}$\thanks{Email: naji.jalil2020@gmail.com}}\\
$^{a,}${\small {\em School of Physics, Damghan University, Damghan, Iran}}\\
$^{b,}${\small {\em Physics Department, Ilam University, Ilam, Iran}}\\
{\small {\em P.O. Box 69315-516, Ilam, Iran}}} \maketitle
\begin{abstract}
\noindent In this paper we consider tachyonic matter in spatially flat FRW universe, and obtain behavior of some important cosmological parameters for two special cases of potentials. First we assume the exponential potential and then consider hyperbolic cosine type potential. In both cases we obtain behavior of the Hubble, deceleration and EoS parameters. Comparison with observational data suggest the model with hyperbolic cosine type scalar-field potentials has good model to describe universe.
{\bf PACs:} 98.80.Jk
\end{abstract}
\section{Introduction}
Recent cosmological observations [1-3] confirmed that our universe expanded while accelerated. Dark energy models come to explain nature of this accelerating expansion. The simplest model to describe the dark energy is the cosmological constant which has two famous problems as the fine-tuning and cosmic coincidence [4]. Hence, another models of the dark energy proposed such as Chaplygin gas model and its extensions [5-12].\\
There are also other interesting models such as quintessence [13-15], phantom [16, 17] and quintom [18-20]. These are based on the scalar fields which plays an important role in cosmology. One of the first major mechanisms where scalar fields are thought to be responsible is the inflationary scenario [21, 22]. Single scalar field is the underlying
dynamics in many inflationary scenario. The mechanism of the initial inflationary and late-time acceleration aspect of our universe may be described by assuming the existence of some gravitationally
coupled scalar fields, with the inflation field generating
inflation and the quintessence field being responsible for
the late accelerated expansion. Differences between models is an effective self-interaction potential. In
the recent work [23], a new formalism for the analysis of
scalar fields in flat isotropic and homogeneous cosmological
models presented. Also, several new accelerating
and decelerating exact cosmological solutions presented based on quintessence scalar field.\\
Recent discoveries [24, 25] have shown a number of novel and unexpected
features, whose explanation will certainly require a deep change in our standard understanding of the universe. Therefore, this paper tries to study arbitrary scalar-field cosmology based on Tachyon cosmological models. The tachyon scalar field was proposed as a source of the dark energy and inflation. The tachyon dark energy has EoS parameter between -1 and 0 [26]. The tachyon is a an unstable field which has became important in string theory through its role in the
Dirac-Born-Infeld action which is used to describe the D-brane action [27].\\
This paper is organized as follows. In next section we review tachyonic matter cosmology and in section 3 we write main equations which govern our model, and should solved for some specific potentials. In section 4 we obtain results corresponding to the exponential potential scalar field. In section 5 we obtain results corresponding to the Hyperbolic cosine type scalar-field. In section 6 we give conclusion.

\section{Tachyonic matter}
Tachyon field described by the following energy density and pressure respectively [28, 29],
\begin{equation}\label{s2}
\rho_{T}=\frac{V(T)}{\sqrt{1-\dot{T}^{2}}}.
\end{equation}
and,
\begin{equation}\label{s1}
p_{T}=-V(T)\sqrt{1-\dot{T}^{2}},
\end{equation}
Therefore, the equation of state (EoS) of tachyon field obtained as follow,
\begin{equation}\label{s3}
\omega_{T}=\frac{p_{T}}{\rho_{T}}=\dot{T}^{2}-1.
\end{equation}
Also, tachyon potential is given by,
\begin{equation}\label{s4}
V(T)=\sqrt{-p_{T}\rho_{T}}.
\end{equation}
The effective action of the tachyon field which
minimally coupled to the gravitational field in the Born-Infeld form is given by,
\begin{equation}\label{s5}
S=\int{d^{4}x\sqrt{-g}\left[\frac{R}{2}+V(T)\sqrt{1+g^{\mu\nu}\partial_{\mu}T\partial_{\nu}T}\right]},
\end{equation}
where $R$ is the curvature scalar, $T$ is the tachyon field, $V(T)$ is
the self-interaction potential, and we used $8\pi G=k=c=1$ units.\\
The tachyonic matter in this model described by a fluid with EoS parameter $-1\leq\omega_{T}\leq0$ where a universe in the accelerating phase smoothly transmit to a phase dominated by a non-relativistic matter.

\section{FRW cosmology}
The spatially flat Friedmann-Robertson-Walker (FRW) Universe is described by the following metric,
\begin{equation}\label{s6}
ds^2=dt^2-a(t)^2(dr^2+r^{2}d\Omega^{2}),
\end{equation}
where $d\Omega^{2}=d\theta^{2}+\sin^{2}\theta d\phi^{2}$. Also, $a(t)$
represents time-dependent scale factor. Therefore, field equations obtained as follows,
\begin{equation}\label{s7}
3H^{2}=\rho_{T},
\end{equation}
and
\begin{equation}\label{s8}
2\dot{H}+3H^{2}=-p_{T},
\end{equation}
where $H=\dot{a}/a$ is Hubble expansion parameter, also $\rho_{T}$ and $p_{T}$ are given by the equations (1) and (2) respectively. Then, the evolution equation for the scalar field is given by [30],
\begin{equation}\label{s9}
\frac{\ddot{T}}{1-\dot{T}^{2}}+3H\dot{T}+\frac{V^{\prime}(T)}{V(T)}=0,
\end{equation}
where the over dot denotes the derivative with respect to the
time-coordinate $t$, while the prime denotes the derivative
with respect to the scalar field $T$, respectively.\\
Combining the equations (1), (2), (7) and (8) we can obtain the following equation describing evolution of Hubble expansion parameter,
\begin{equation}\label{s10}
\dot{H}=\frac{V^{2}(T)}{6H^{2}}-\frac{3H^{2}}{2}.
\end{equation}
Also, combining the equations (1), (7) and (9) we can obtain the following equation describing evolution of the scalar field,
\begin{equation}\label{s11}
\frac{\ddot{T}}{1-\dot{T}^{2}}+\sqrt{\frac{3V(T)}{\sqrt{1-\dot{T}^{2}}}}\dot{T}+\frac{dV(T)}{dT}V(T)^{-1}=0.
\end{equation}
The deceleration parameter $q$ is an important observational quantity which is given by,
\begin{equation}\label{s12}
q=-\frac{\dot{H}}{H^{2}}-1=\frac{\dot{T}^{2}}{2}-1.
\end{equation}
In the case of constant $T$ we yield $q=-1$ which is de Sitter type accelerated universe.\\
Now, we try to solve the above equations and extract cosmological parameters for some special cases of exponential and hyperbolic potentials.

\section{The exponential potential scalar field}
\begin{figure}[h!]
 \begin{center}$
 \begin{array}{cccc}
\includegraphics[width=50 mm]{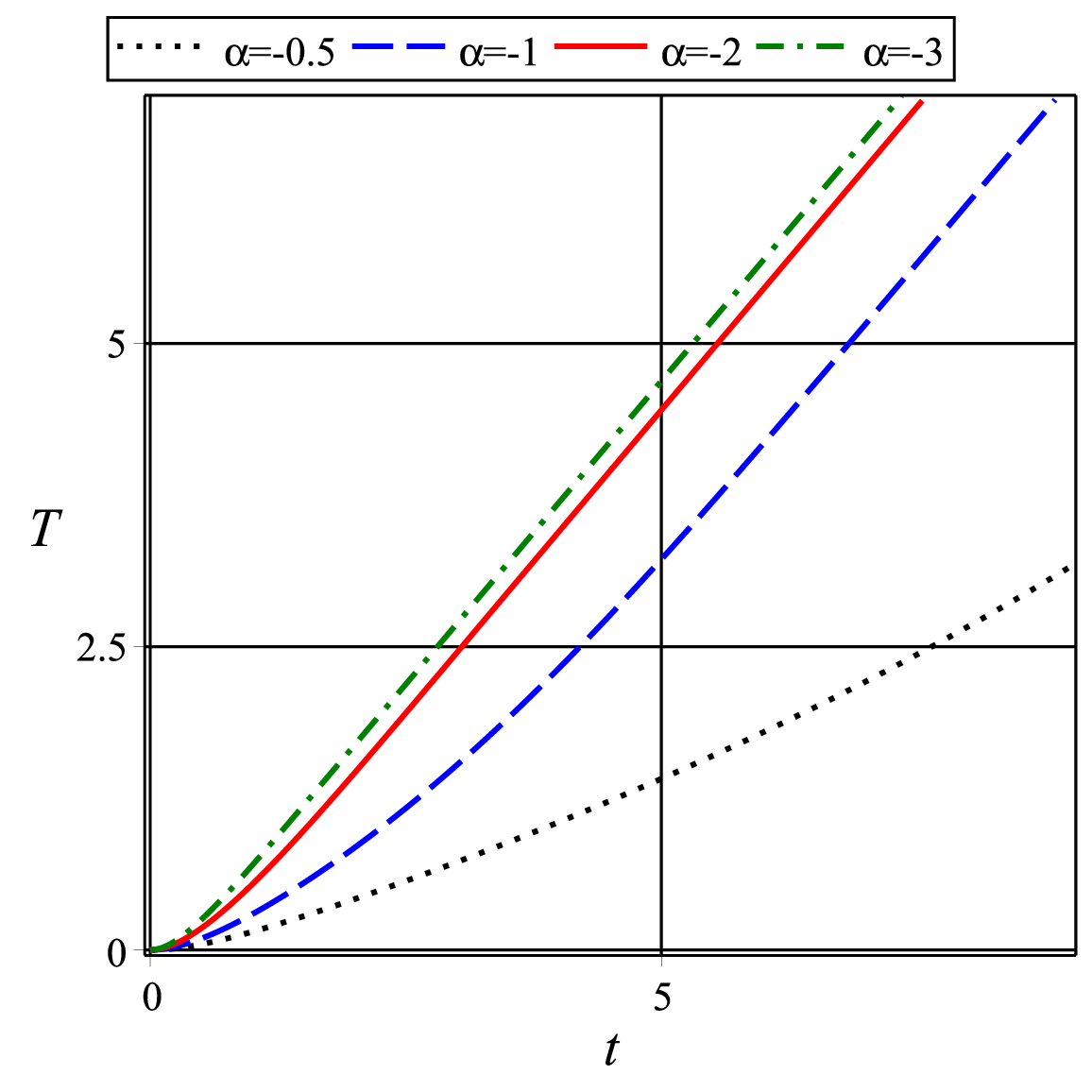}
 \end{array}$
 \end{center}
\caption{Tachyon field in terms of $t$ for $V_{0}=1$. $\alpha=-0.5$ (Black dotted), $\alpha=-1$ (blue dashed), $\alpha=-2$ (red solid), $\alpha=-3$ (green dash dotted) for initial values $T(0)=\dot{T}(0)=0$.}
 \label{fig1}
\end{figure}

The number of known exact solutions for
cosmological models based on scalar fields is
rather limited. One of such models is the flat Friedmann
universe filled with a minimally coupled scalar
field with exponential potential. This solution describes a power-law expansion
of the universe. The tachyon potential is given by [31],
\begin{equation}\label{s13}
V=V_{0}e^{\alpha T},
\end{equation}
where $V_0$ and $\alpha$ (tachyon mass) are arbitrary constants. It order to obey $V\rightarrow0$ at $T\rightarrow\infty$ we need to choose negative $\alpha$. In that case we have a condition as,
\begin{equation}\label{s14}
\frac{V^{\prime}}{V}=const.
\end{equation}
These types of exponential potential are important in four-dimensional
effective Kaluza-Klein type theories from compactification
of the higher-dimensional supergravity or superstring theories [32] and may arise due to non-perturbative effects such as gaugino condensation [33]. In that case, the equation (11) rewritten as the following form,
\begin{equation}\label{s15}
\frac{\ddot{T}}{1-\dot{T}^{2}}+\sqrt{\frac{3V_{0}e^{\alpha T}}{\sqrt{1-\dot{T}^{2}}}}\dot{T}+\alpha=0.
\end{equation}
In the Ref. [30], above equation considered and behavior of $T$ in terms of $\dot{T}$ obtained to see attractor behavior which is important in inflationary scenario. Here we would like to obtain evolution of tachyon field and try to obtain time dependent $T$.
Numerically, we find time evolution of the tachyon field in the Fig. \ref{fig1}. We can see that it is increasing function of time. Also, it is find that increasing $|\alpha|$ increases value of the tachyon field.\\

\begin{figure}[h!]
 \begin{center}$
 \begin{array}{cccc}
\includegraphics[width=50 mm]{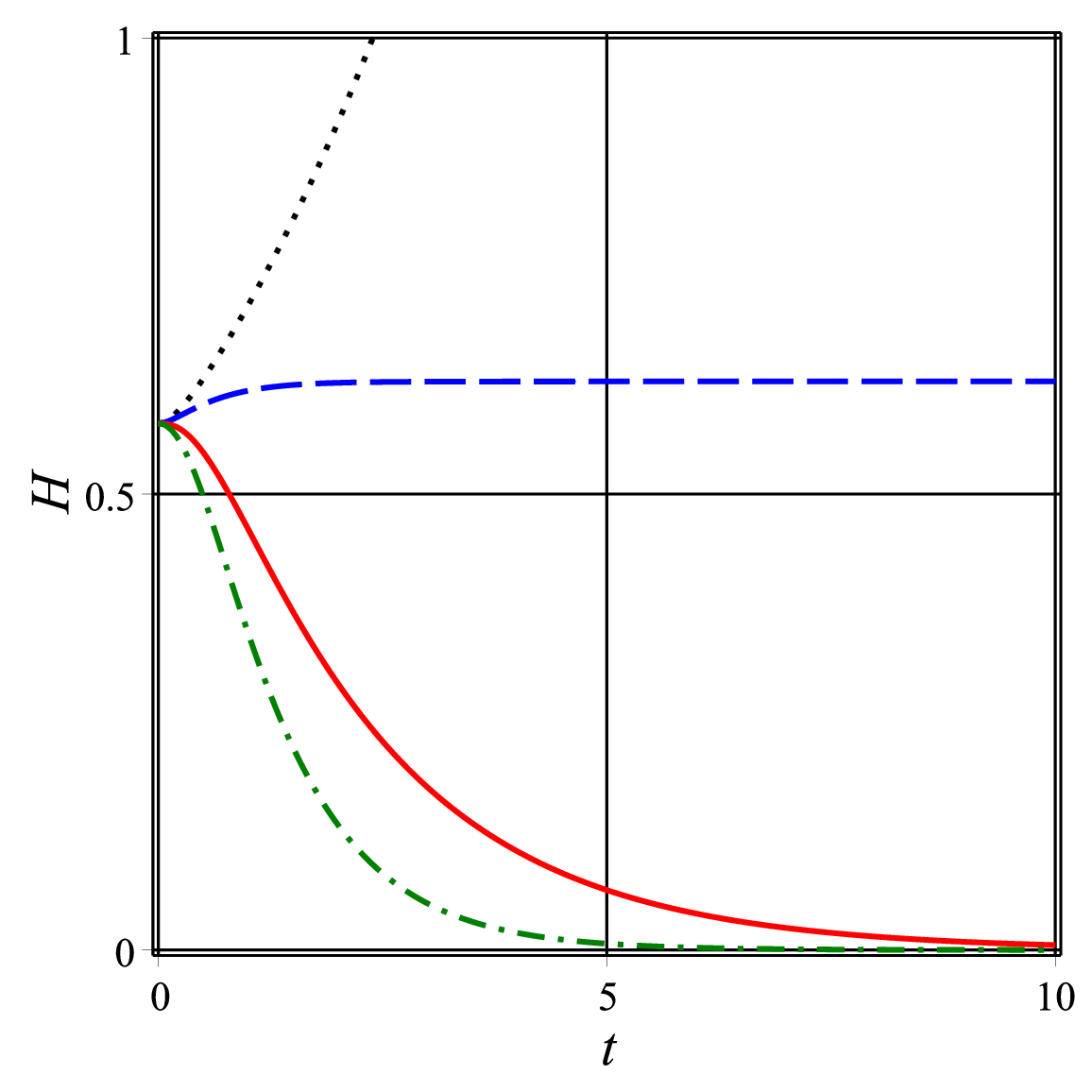}
 \end{array}$
 \end{center}
\caption{Hubble expansion parameter in terms of $t$ for $V_{0}=1$, $A=1$, $B=-0.5$, $C=0.5$ and $\beta=2$. $\alpha=-0.5$ (Black dotted), $\alpha=-1$ (blue dashed), $\alpha=-2$ (red solid), $\alpha=-3$ (green dash dotted)}
 \label{fig2}
\end{figure}

It is general behavior of the tachyon field. In order to obtain other important cosmological parameter we need an analytical expression of the tachyon field.\\
We suggest the following form of time dependent tachyon field,
\begin{equation}\label{s16}
T=At+B+Ce^{-\beta t},
\end{equation}
where $A$, $B$, $C$ and $\beta$ are arbitrary constant. We can fix these constant to obtain behavior corresponding to Fig. \ref{fig1}. For example choosing $B=-0.5$, $C=0.5$, $\beta=2$ and $A=1$ or $A=0.4$ give exactly solid red or dotted lack lines of the Fig. \ref{fig1}, respectively.\\
By using the equations (1), (7) and (\ref{s16}) we can obtain Hubble expansion parameter as follow,
\begin{equation}\label{s16-1}
H=\frac{\sqrt{3}}{3}\sqrt{\frac{V_{0}e^{\alpha(At+B+Ce^{-\beta t})}}{\sqrt{1-(A-C\beta e^{-\beta t})^{2}}}}.
\end{equation}
In the Fig. \ref{fig2} we can see behavior of the Hubble expansion parameter in terms of cosmic time. We can see that $|\alpha|>1$ yields to reasonable result where Hubble expansion parameter yields to a constant at the late time as expected. It is illustrated that the value of the Hubble parameter decreased by $\alpha$.

Using the approximation given by the equation (\ref{s16}) and equation (12) we can obtain other cosmological parameters such as the deceleration parameter,
\begin{equation}\label{s17}
q=-1+\frac{1}{2}\left(A-C\beta e^{-\beta t}\right)^{2}.
\end{equation}

\begin{figure}[h!]
 \begin{center}$
 \begin{array}{cccc}
\includegraphics[width=50 mm]{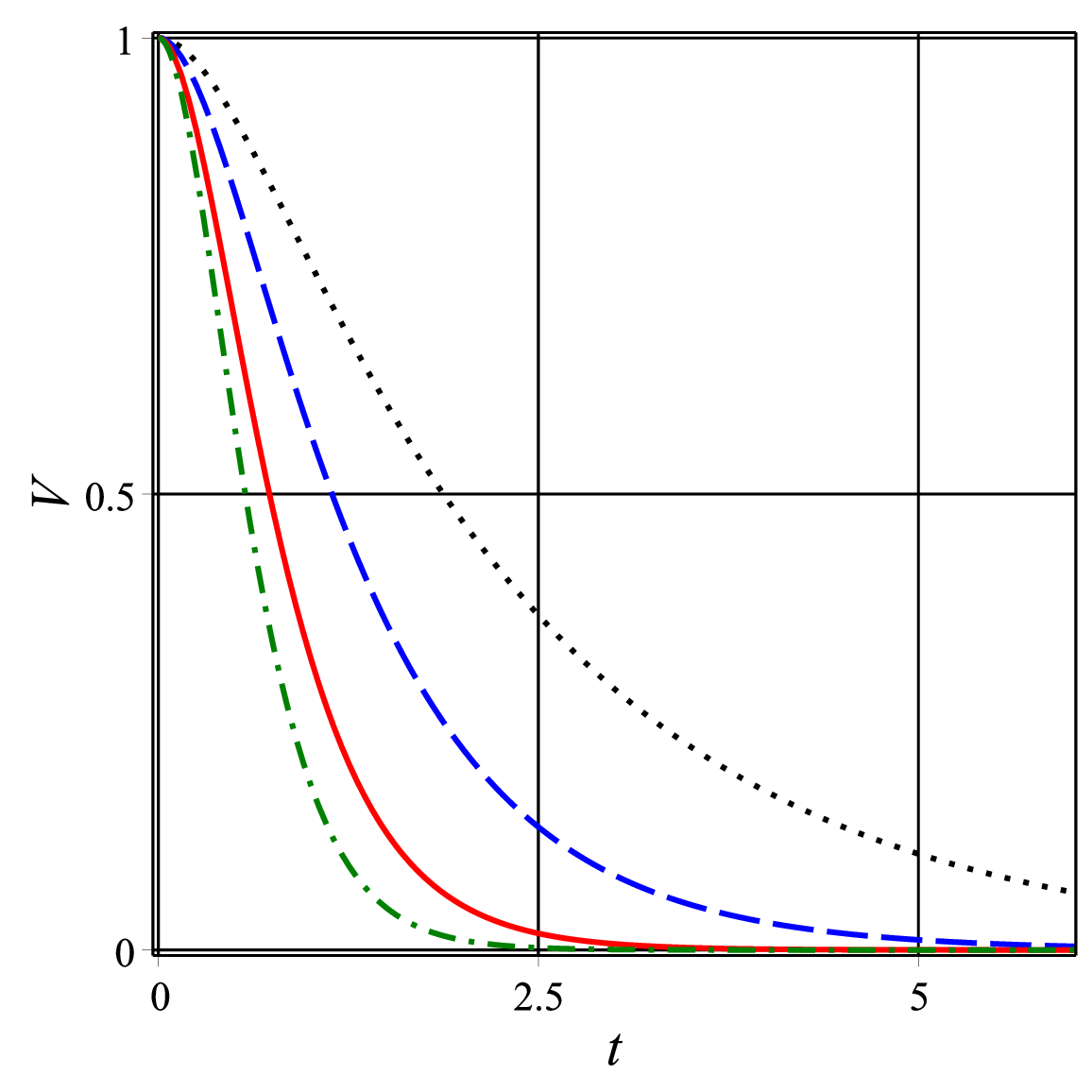}
 \end{array}$
 \end{center}
\caption{Tachyon potential in terms of $t$ for $V_{0}=1$, $A=1$, $B=-0.5$, $C=0.5$ and $\beta=2$. $\alpha=-0.5$ (Black dotted), $\alpha=-1$ (blue dashed), $\alpha=-2$ (red solid), $\alpha=-3$ (green dash dotted).}
 \label{fig3}
\end{figure}

It is easy to find that (selecting constant like $C=0.5$, $\beta=2$ and $A=0.4$) the deceleration parameter is decreasing function of time which yields to -1 at the late time. Corresponding to the value of the constant $C$ ($|C|>1$), it is possible to see accelerating to decelerating phase transition.\\
Also using the equation (3), we can obtain equation of state parameter as the follow,
\begin{equation}\label{s18}
\omega=-1+\left(A-C\beta e^{-\beta t}\right)^{2},
\end{equation}
which guarantee that $\omega\geq-1$ for all possible choice of constant. We can fix constant as before to see that the EoS parameter yields to -1 at the late time.\\
Finally, we can see behavior of the tachyon potential in the Fig. \ref{fig3}. For the very small values of parameter $|\alpha|$ the tachyon potential behaves as a constant. Generally we have a maximum for the potential and it yields to zero at the late time.

\section{Hyperbolic cosine type potential}
It is also possible to consider the hyperbolic cosine type potential given by,
\begin{equation}\label{s19}
V=V_{0}\cosh^{\nu}[\gamma(T-T_{0})],
\end{equation}
where $V_0$, $\nu$, $\gamma$ and $T_{0}$ are arbitrary constants. $T_{0}$ is initial value of tachyon field which may be considered zero as before, however we assume small initial value to avoid divergency of solution. In that case the condition (14) is no longer valid.
Therefore, the equation (11) rewritten as the following form,
\begin{equation}\label{s20}
\frac{\ddot{T}}{1-\dot{T}^{2}}+\sqrt{\frac{3V_{0}\cosh^{\nu}[\gamma(T-T_{0})]}{\sqrt{1-\dot{T}^{2}}}}\dot{T}+\frac{dV(T)}{dT}V(T)^{-1}=0.
\end{equation}
General solution of the equation (\ref{s20}) presented in Fig. 4 numerically. Similar to the previous case we can see that the scalar field is increasing function of time for negative $\nu$. It is illustrated that the scalar field yields to a constant at the late time for $\nu=0$. 

\begin{figure}[h!]
 \begin{center}$
 \begin{array}{cccc}
\includegraphics[width=50 mm]{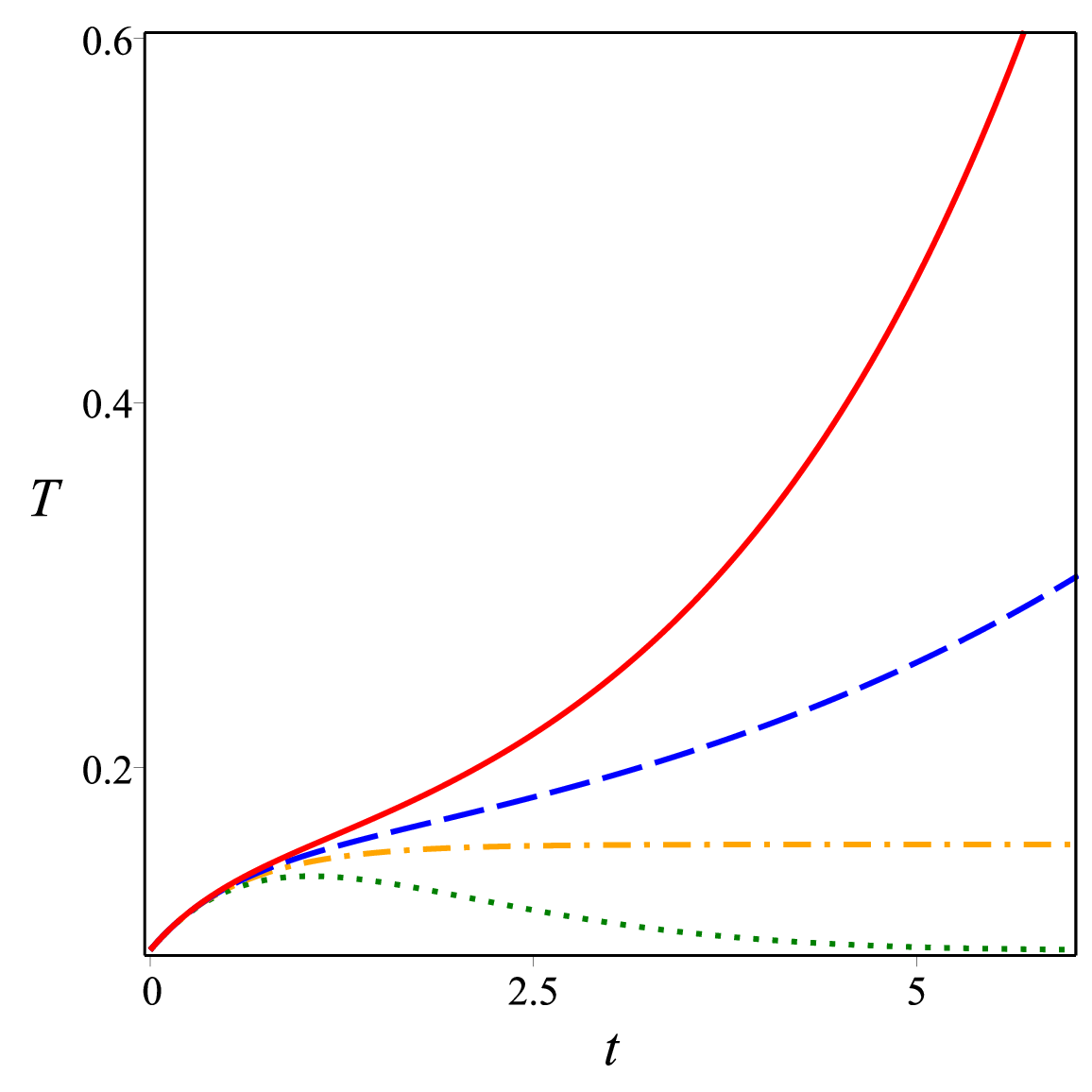}
 \end{array}$
 \end{center}
\caption{Tachyon field in terms of $t$ for $T_{0}=0.1$, $V_{0}=1$, and $\gamma=1$. $\nu=1$ (dotted green), $\nu=0$ (dashed orange), $\nu=-0.5$ (dashed blue), $\nu=-1$ (solid red).}
 \label{fig4}
\end{figure}

\begin{figure}[h!]
 \begin{center}$
 \begin{array}{cccc}
\includegraphics[width=50 mm]{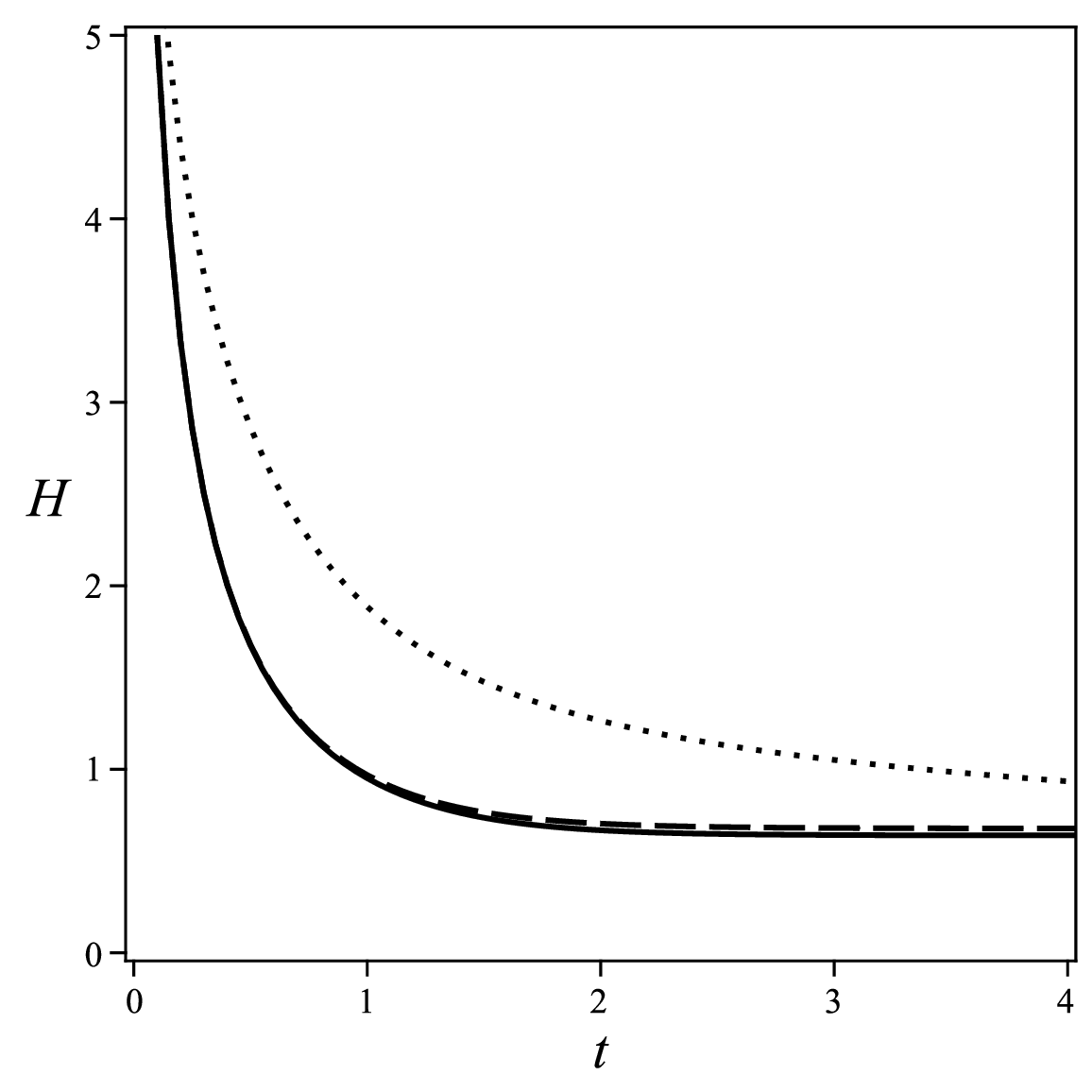}
 \end{array}$
 \end{center}
\caption{Hubble expansion parameter in terms of $t$ for $V_{0}=1$, $C_{1}=5$, $C_{2}=5$, $\gamma=0.5$ (dotted line), $\gamma=0.1$ (dashed line), $\gamma=0.01$ (solid line).}
 \label{fig5}
\end{figure}

If we assume $\dot{\phi}\ll1$ and $\gamma\ll1$, then the explicit form of the scalar field will be available,
\begin{equation}\label{s21}
T=C_{1}e^{-\frac{\sqrt{3}}{6}(3\sqrt{V_{0}}-\sqrt{9V_{0}-12V_{0}\gamma^{2}})t}+C_{2}e^{-\frac{\sqrt{3}}{6}(3\sqrt{V_{0}}+\sqrt{9V_{0}-12V_{0}\gamma^{2}})t},
\end{equation}
where $C_{1}$ and $C_{2}$ are integration constants. Using this approximation we can obtain other cosmological parameters such as the deceleration parameter.
It is easy to find that the deceleration parameter is decreasing function of time yields to -1 at the late time similar to the previous model. It is also possible to see accelerating to decelerating phase transition. Also we can obtain equation of state parameter and find that $\omega\geq-1$. It is clear that the EoS parameter yields to -1 at the late time.\\
We can perform numerical analysis on the Hubble expansion parameter. In the Fig. \ref{fig5} we draw Hubble expansion parameter versus cosmic time. We can see that it is decreasing function of time which yields to a constant value at the late time which is expected. It is illustrated that the value of the Hubble parameter decreased by $\gamma$.\\

\begin{figure}[h!]
 \begin{center}$
 \begin{array}{cccc}
\includegraphics[width=50 mm]{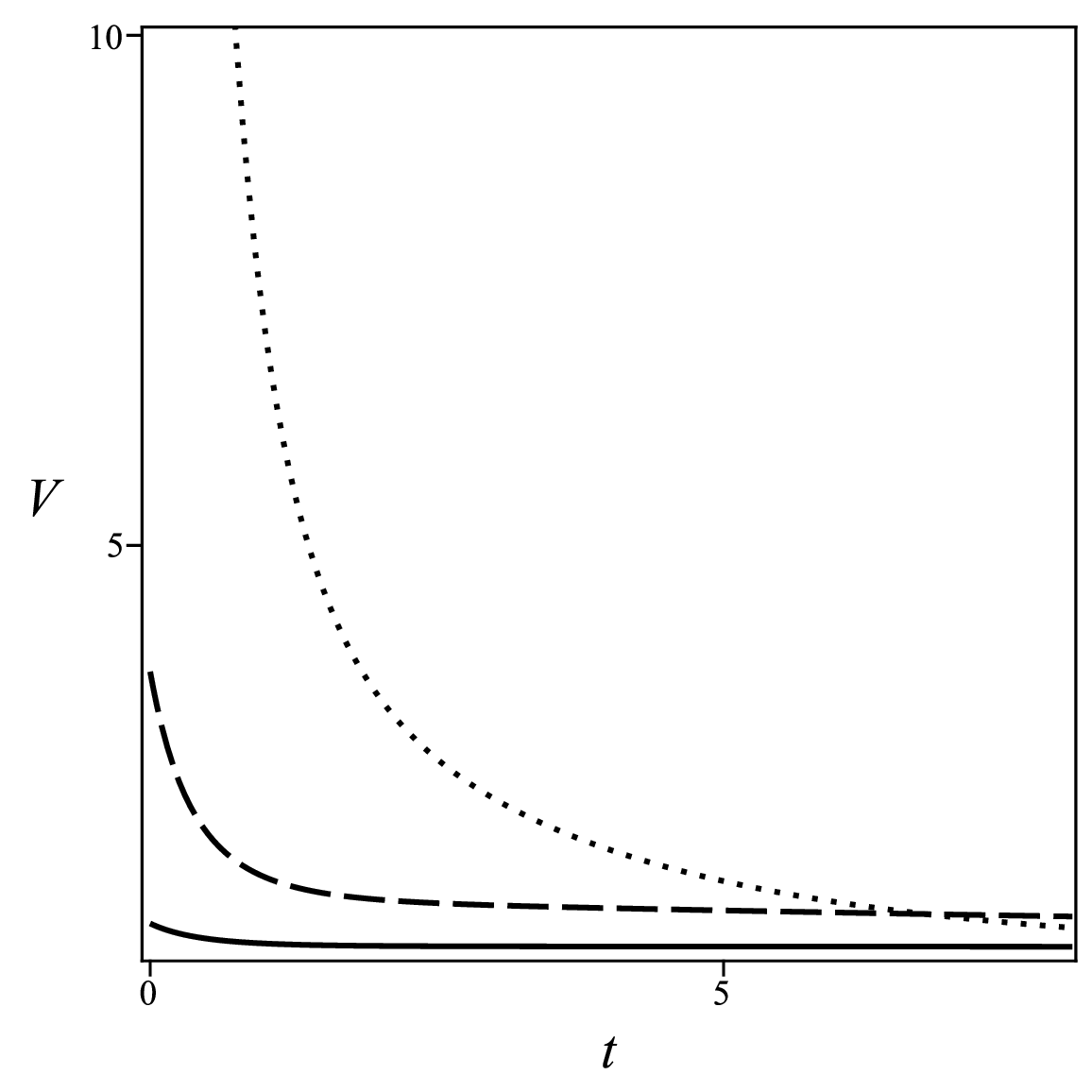}
 \end{array}$
 \end{center}
\caption{Potential in terms of $t$ for $V_{0}=1$, $C_{1}=5$, $C_{2}=5$, $\gamma=0.5$ (dotted line), $\gamma=0.2$ (dashed line), $\gamma=0.075$ (solid line).}
 \label{fig6}
\end{figure}

Finally, we can see behavior of the scalar potential in the Fig. \ref{fig6}. For the very small values of parameter $\gamma$ the scalar potential behaves as a constant. Generally, the potential is decreasing function of time similar the previous model.\\
We found reasonable behavior of our models, however we have to add ordinary matter to the models and compare our results with appropriate observational data such as $H(z)$ data [34].

\section{Matter contribution}
In order to have comparison with observational data we should consider all matter contribution in our  model. In that case we have the following conservation equation,
\begin{equation}\label{s22}
\dot{\rho}+3H(p+\rho)=0,
\end{equation}
where $\rho=\rho_{\phi}+\rho_{m}$ and $p=p_{\phi}+p_{m}$, with matter density $\rho_{m}$ and pressure $p_{m}$. Now, the equations (7) and (8) extended to the following relations,
\begin{equation}\label{s23}
3H^{2}=\rho,
\end{equation}
and,
\begin{equation}\label{s24}
2\dot{H}=-p-\rho.
\end{equation}
We assume non-interacting case, therefore conservation equation (\ref{s22}) separates as follow,
\begin{equation}\label{s25}
\dot{\rho}_{T}+3H(p_{T}+\rho_{T})=0,
\end{equation}
and
\begin{equation}\label{s26}
\dot{\rho}_{m}+3H(p_{m}+\rho_{m})=0,
\end{equation}
with $\omega_{m}=p_{m}/\rho_{m}$ as EoS of matter. An important parameter to compare with observational data is $H(z)$. We will calculate $H(z)$ for two different cases of exponential and Hyperbolic scalar potential.
\subsection{Exponential potential}
In this case we use relation (13) and investigate behavior of Hubble expansion parameter versus redshift. We can see good agreement with observational data for $0.01\leq|\alpha|\leq0.5$. However, $0.1<|\alpha|<0.5$ yields to the best agreement with high redshift at the early universe.

\begin{figure}[h!]
 \begin{center}$
 \begin{array}{cccc}
\includegraphics[width=50 mm]{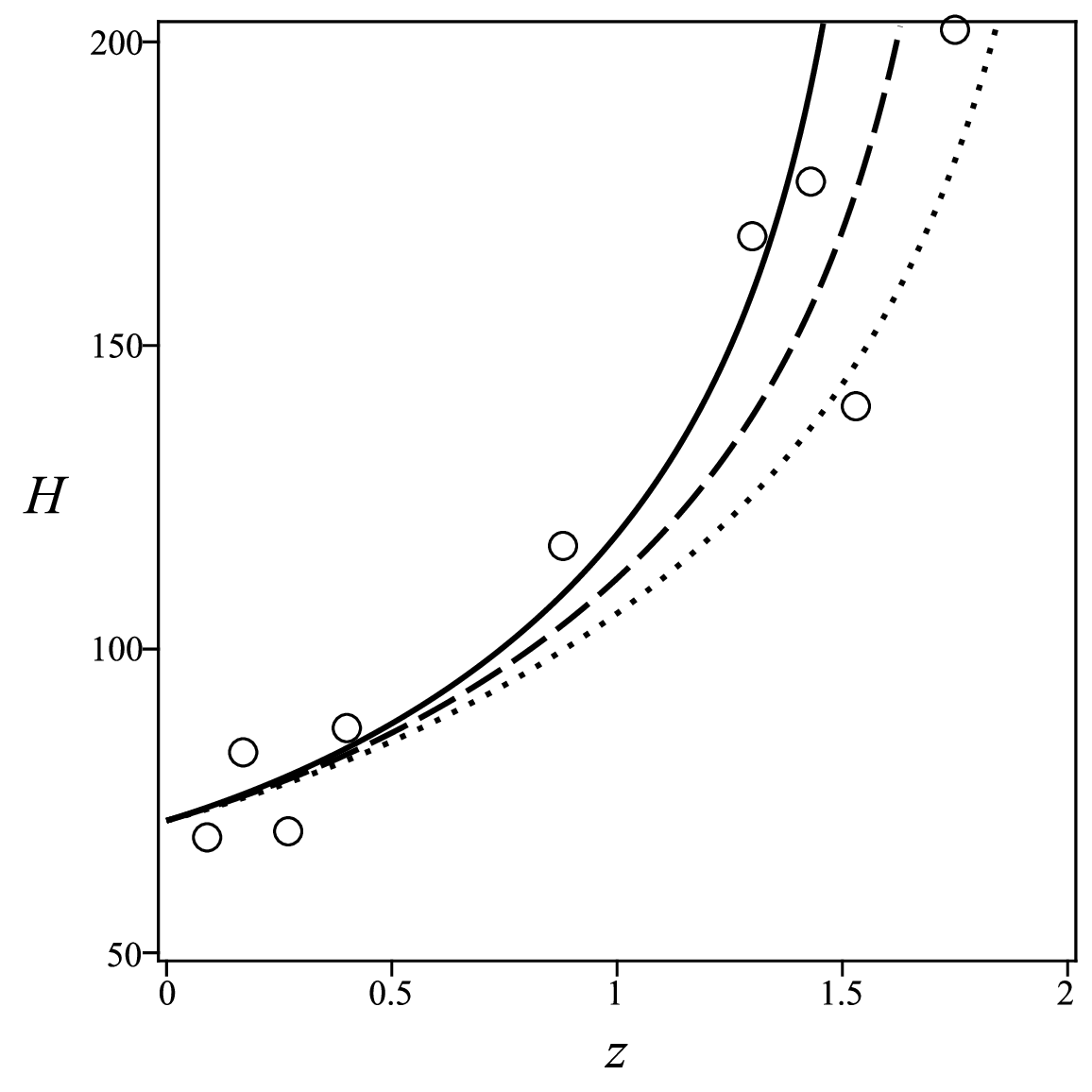}
 \end{array}$
 \end{center}
\caption{Hubble expansion parameter in terms of $z$ for $V_{0}=1$, $C=5$, $B=5$, $\alpha=-0.5$ (dotted line), $\alpha=-0.1$ (dashed line), $\alpha=-0.01$ (solid line). Big dots denote observational data.}
 \label{fig7}
\end{figure}

\subsection{Hyperbolic potential}
In this case we use relation (\ref{s19}) and investigate behavior of Hubble expansion parameter versus redshift. We can see good agreement with observational data for $0.1\leq\gamma\leq0.5$ and $\nu=-1$.\\
Comparison of both models with each other suggest that the second model (Hyperbolic potential) is more appropriate.

\begin{figure}[h!]
 \begin{center}$
 \begin{array}{cccc}
\includegraphics[width=50 mm]{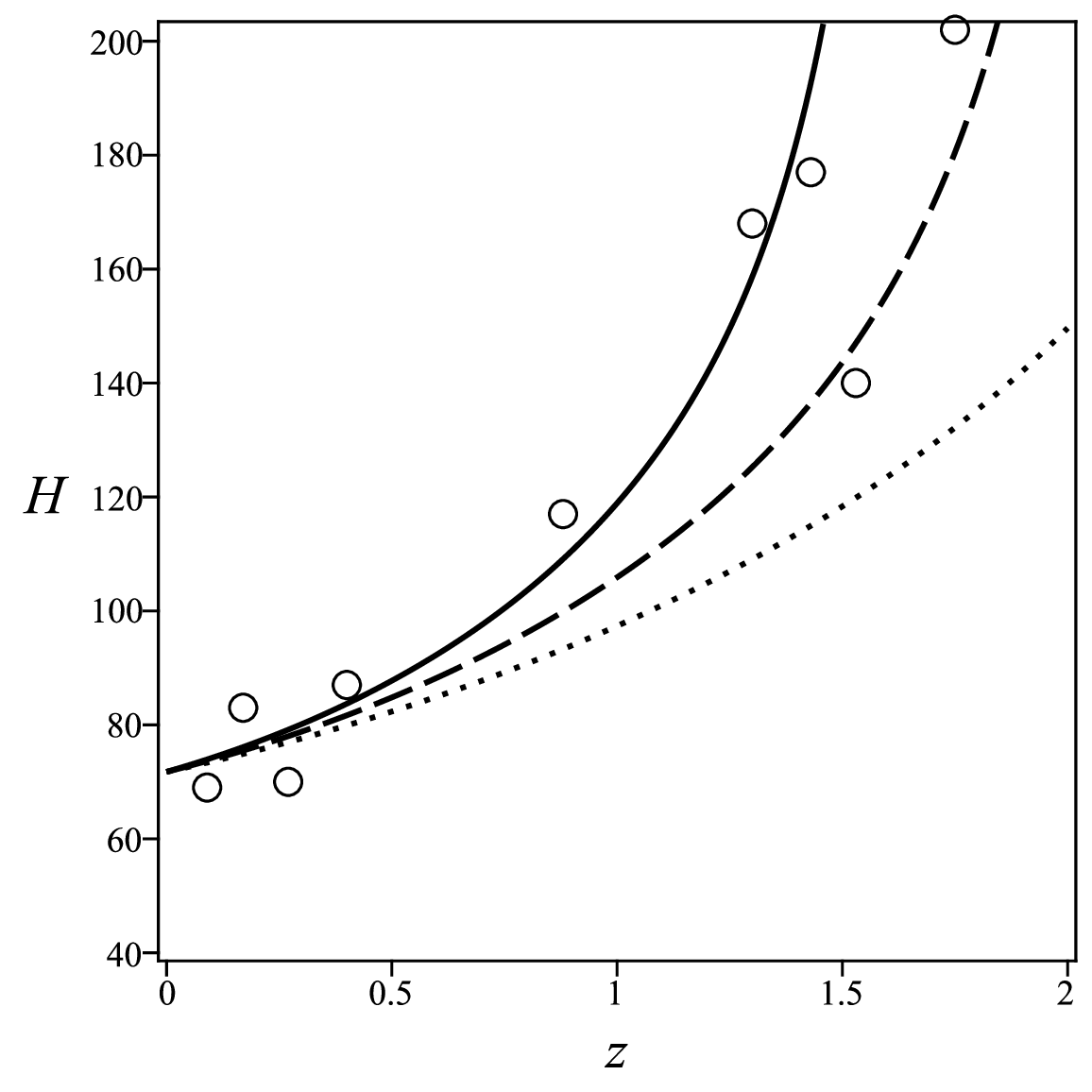}
 \end{array}$
 \end{center}
\caption{Hubble expansion parameter in terms of $z$ for $V_{0}=1$, $C_{1}=5$, $C_{2}=5$, $\gamma=0.5$ (dotted line), $\gamma=0.1$ (dashed line), $\gamma=0.01$ (solid line). Big dots denote observational data.}
 \label{fig8}
\end{figure}
\section{Conclusion}
In this work, we considered tachyon scalar field in flat FRW universe and proposed two different models based on tachyon field potential. In the first model, we assumed the exponential potential scalar field and solved the equation describing the evolution of the tachyon field numerically. We found that the tachyon field in this model has totally positive value and increased by time. It means that at the late time the scalar field takes infinite value. In order to obtain other cosmological parameters we fit tachyon field and found explicit expression for that, then deceleration and EoS parameters obtained. Also, we discussed numerically about tachyon potential and Hubble expansion parameter. We found that the deceleration parameter as well as EoS parameter is decreasing function of time and yields to -1 at the late time. We have shown that the Hubble expansion parameter is decreasing function of time and yields to a constant value at the late time as expected. Finally we found corresponding to the very small values
of parameter $\alpha$ the scalar potential behaves as a constant. In general, there is a maximum
for the potential and it yields to zero at the late time. Apart of the unlike behavior of the tachyon field for inconsistent model parameters, the model yields to good results. It yields us to investigate another model to obtain also good behavior for the tachyon field. In the second model the hyperbolic cosine type potential considered. Our numerical analysis have shown that the tachyon field is increasing function of time with the positive value. It is illustrated that the tachyon field may yields to a constant at the late time. It is more suitable than the first model and yields to good behavior cosmological parameters like Hubble, deceleration and EoS parameters. Therefore, we can suggest the second model as a good model to describe universe. In order to confirm our claim we added other matter to the model obtained Hubble expansion parameter in terms of redshift and compare our results with observational data to find that the second model is in agreement with observational data.\\

\end{document}